\documentclass{article}
\usepackage{amsmath,amssymb,amsfonts,graphicx,tikz,hyperref,bm,cprotect}
\usepackage[left=2cm,right=2cm]{geometry}
\usetikzlibrary{arrows}
\author{Shikhar Mittal\footnote{shikhar.mittal4@gmail.com, shikhar.mittal17@imperial.ac.uk}}
\title{Reflection of a Point Object in an Arbitrary Curved Mirror}
\newcommand{\ud}{\mathrm{d}}
\hypersetup{linkcolor=blue, colorlinks=true} 
\makeatletter
\newsavebox\myboxA
\newsavebox\myboxB
\newlength\mylenA

\newcommand*\xoverline[2][0.75]{%
    \sbox{\myboxA}{$\m@th#2$}%
    \setbox\myboxB\null
    \ht\myboxB=\ht\myboxA%
    \dp\myboxB=\dp\myboxA%
    \wd\myboxB=#1\wd\myboxA
    \sbox\myboxB{$\m@th\overline{\copy\myboxB}$}
    \setlength\mylenA{\the\wd\myboxA}
    \addtolength\mylenA{-\the\wd\myboxB}%
    \ifdim\wd\myboxB<\wd\myboxA%
       \rlap{\hskip 0.5\mylenA\usebox\myboxB}{\usebox\myboxA}%
    \else
        \hskip -0.5\mylenA\rlap{\usebox\myboxA}{\hskip 0.5\mylenA\usebox\myboxB}%
    \fi}
\makeatother
\begin{document}
\maketitle
\begin{abstract}
In this work, I have derived the equation of the curve obtained on reflection of a point object in an arbitrary curved mirror if the object and the mirror are placed on the 2D Cartesian plane. I have used only the basic laws of reflection of classical geometric optics and elementary coordinate geometry. Several examples are provided and compared with Gaussian optics. We also see how the equations reduce to the standard mirror formula under the paraxial approximation.
\end{abstract}
\section{Introduction}
\noindent The study of light is divided into two categories \cite{optics}: 1) Physical optics -- which is concerned with the nature of light or more generally the electromagnetic radiation and its interaction with matter. 2) Geometric optics -- is the study of the phenomenon of refraction and reflection. This work deals with the latter, particularly image formation in randomly shaped mirrors. The formation of images is described by the following empirical laws of reflection:
\begin{enumerate}
    \item The incident ray, normal and the reflected ray lie in the same plane.
    \item The angle of incidence $\angle i$ is equal to the angle of reflection $\angle r$.
\end{enumerate}
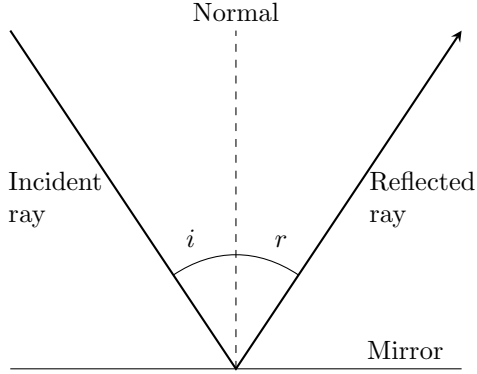
\begin{figure}[t]
\centering
\begin{tikzpicture}[scale=1.5,>=stealth]
\draw (-2,0)--(2,0);
\node[above] at (1.5,0){Mirror};
\draw[dashed] (0,0)--(0,3);
\node[above] at (0,3){Normal};
\draw [thick,->] (-2,3)--(0,0)--(2,3);
\node[right, align=left] at (1.1,1.5){Reflected\\ ray};
\node[left, align=left] at (-1.1,1.5){Incident\\ ray};
\node[above] at (0.4,1){$r$};
\node[above] at (-0.4,1){$i$};
\draw (0.5547,0.832) to [out=146.31, in= 33.69] (-0.5547,0.832);
\end{tikzpicture}
\caption{Laws of reflection}
\end{figure}
In the context of formation of images in mirrors or lenses, the following key assumptions are used which collectively form the Gaussian optics \cite{gerrard}:
\begin{enumerate}
    \item If the wavelength of light is small, then the propagation of light can be described by individual light rays rather than the Huygen's wavefronts. A Gaussian beam\footnote{A Gaussian beam is a monochromatic beam of light (electromagnetic wave) whose magnetic and electric field amplitude profiles are given by the Gaussian function \cite{gerrard}.} is a good approximation for a single light ray.
    \item Paraxial approximation -- when the light rays are assumed to lay close to and make only small angles with the optical axis of the system. This allows us to use small angle approximations $\sin\theta\approx\theta$ and $\tan\theta\approx\theta$ which are correct to first order for small $\theta$.
\end{enumerate}
For spherical mirrors, the paraxial approximation is possible for mirrors of small aperture and small objects. Under these conditions a simple relation can be derived between the image and object distance with the radius of curvature of the mirror \cite{optics}. The following is known as the mirror formula,
\begin{equation}
  \frac{1}{v}+\frac{1}{u}=\frac{1}{F},\label{Mirfor}
\end{equation}
where $u,v$ and $F$ are object distance, image distance and focal length of the mirror, respectively. The focal length $F$ is related to the radius of curvature $R$ as$$F=\frac{R}{2}.$$The physical interpretation of focus is that if parallel beam of light rays fall on the mirror they converge to a unique point called the focus. If the assumptions of Gaussian optics are relaxed then the image obtained in general is distorted. This phenomenon is known as \textit{aberration}. Aberration is of several types, but the one of interest here is the so-called \textit{spherical aberration}. This occurs when the parallel beam of light has a large cross section and they do not focus at a single point \cite{lipson}. We want to find out how the image looks like without the paraxial approximation and if the mirror is in any random shape and not just simple ones like plane or spherical ones. Wherever applicable, I have used \verb|MATLAB R2018a Update 3 (9.4.0.885841)| for the purpose of plotting and numerical solution of an equation. Before we go to the main calculations, we need to introduce a small concept from coordinate geometry.
\subsection{Locus of intersection of family of lines}\label{Sec1}
\noindent A family of lines is given by $$\mathcal{L}(\lambda)\equiv y=m(\lambda)x+c(\lambda),$$for some parameter $\lambda\in\mathbb{R}$. Our aim is to find the locus of intersection point of this set. Consider the intersection of the two lines that are infinitesimally apart, $\mathcal{L}(\lambda)$ and $\mathcal{L}(\lambda+\ud\lambda)$, and suppose that the lines meet at $(h,k)$.
\begin{align}
k&=mh+c,\label{E1}\\
k&=(m+\ud m)h+(c+\ud c),\label{E2}
\end{align}
where
\begin{align*}
\ud m& = \frac{\ud m}{\ud\lambda}\ud\lambda,\\
\ud c& = \frac{\ud c}{\ud\lambda}\ud\lambda.
\end{align*}
Subtract Eq. \eqref{E1} from Eq. \eqref{E2} to get
\begin{equation}
h=-\frac{\ud c}{\ud m}.
\end{equation}
Put this back into Eq. \eqref{E1} to get
\begin{equation}
k=c-m\frac{\ud c}{\ud m}.
\end{equation}
So that the locus is given by
\begin{align}
&\bm{x(\lambda)=-\frac{c'}{m'}}\label{E3}\\
&\bm{y(\lambda)=c-m\frac{c'}{m'}},\label{E4}
\end{align}
where prime $(')$ denotes differentiation with respect to $\lambda$. The parameter $\lambda$ may be eliminated to obtain an equation in $x$ and $y$.
\subsubsection*{An example}
\begin{figure}[t]
    \centering
    \includegraphics[width=1\linewidth]{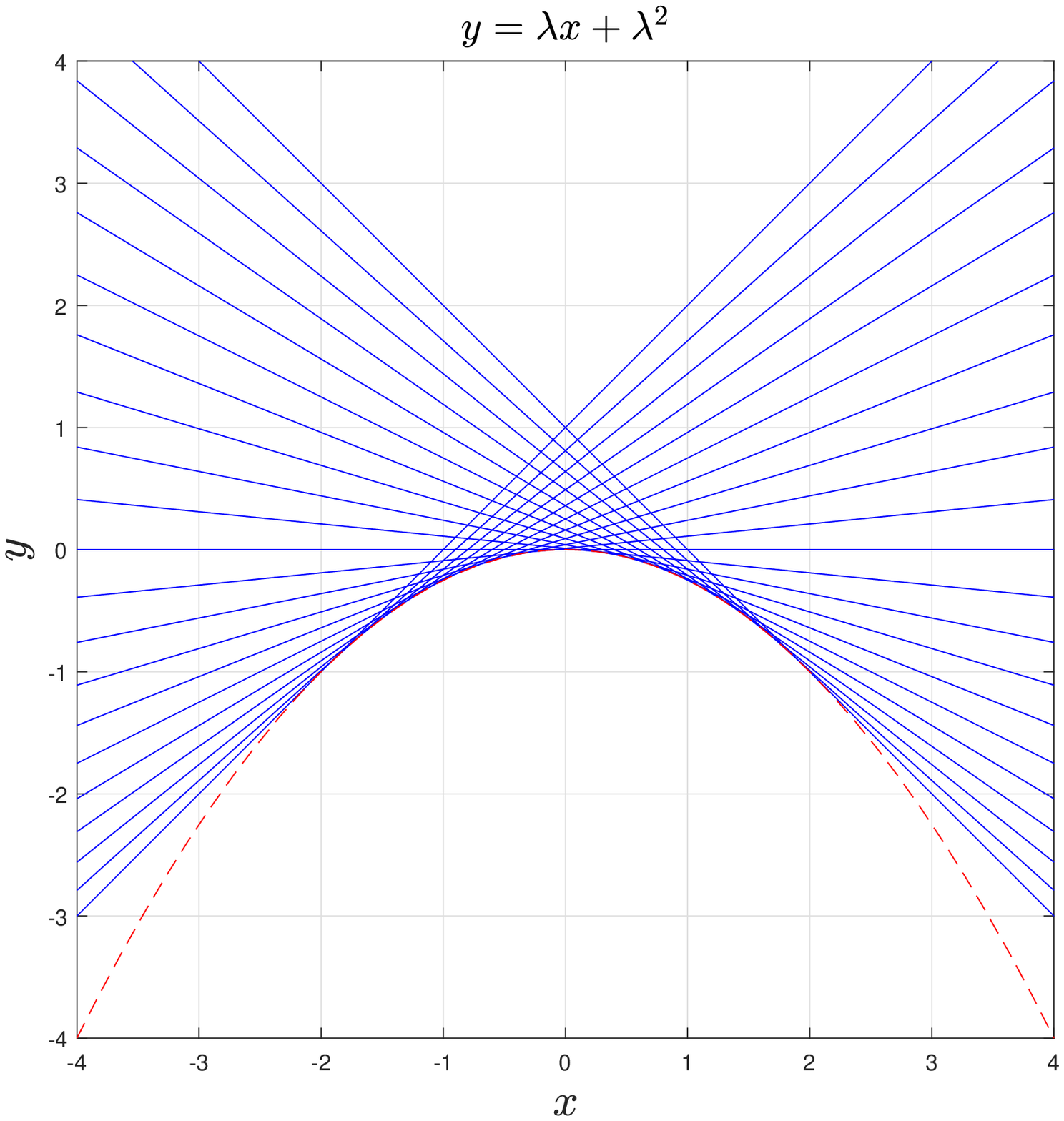}
    \caption{The red dashed line is the curve $x^2+4y=0$. The blue lines represent the family: $y=\lambda x+\lambda^2$ for $\lambda\in[-2,2]$. Clearly, in this range the intersection of blue lines and the red curve overlap exactly.}
    \label{fig:Fig1}
\end{figure}

\noindent Consider the following example; find the locus of intersection of the family of lines given by$$y=\lambda x+\lambda^2.$$For this set,
\begin{align*}
m(\lambda)=\lambda,\\
c(\lambda)=\lambda^2.
\end{align*}
Using Eq. \eqref{E3} and \eqref{E4}, the locus is given by
\begin{align*}
x(\lambda)=-2\lambda,\\
y(\lambda)=-\lambda^2.
\end{align*}
On eliminating $\lambda$ the equation obtained is$$x^2+4y=0.$$See Fig. \ref{fig:Fig1} for a graphical illustration of this example.
\section{Reflection in a Mirror Given by an Equation}
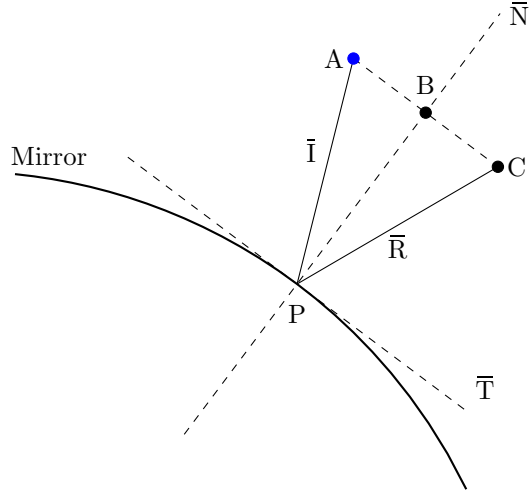
\begin{figure}[t]
\centering
\begin{tikzpicture}[scale=1.5,>=stealth]
\draw[black, thick, domain=0.5:4.5] plot (\x, {sqrt(25-\x*\x)});
\node [below] at (3,3.9){P};
\draw[black,dashed,domain=2:4.8] plot(\x,{4*\x/3});
\draw[dashed,domain=1.5:4.5] plot(\x,{-3*\x/4+25/4});
\draw[dashed](3.5,6)--(4.78,5.04);
\draw(3,4)--(4.78,5.04);
\draw(3,4)--(3.5,6);
\draw [fill, blue] (3.5,6) circle [radius=0.05];
\draw [fill, black] (4.78,5.04) circle [radius=0.05];
\draw [fill, black] (4.14,5.52) circle [radius=0.05];
\node [above] at (4.14,5.6){B};
\node [right] at (4.78,5.04){C};
\node [left] at (3.5,6){A};
\node [below] at (0.8,5.3){Mirror};
\node [right] at (4.8,6.4){$\xoverline{\text{N}}$};
\node [above left] at (3.25,5){$\xoverline{\text{I}}$};
\node [below] at (3.89,4.52){$\xoverline{\text{R}}$};
\node [above right] at (4.5,2.9){$\xoverline{\text{T}}$};
\end{tikzpicture}\label{Fig1}
\caption{The blue dot represents the given point A. $\xoverline{\text{I}}$ is the incident ray which is incident on the mirror at P. Reflected ray $\xoverline{\text{R}}$ passes through C which is symmetrical to A about the normal. The tangent and normal to the mirror at P are represented by $\xoverline{\text{T}}$ and $\xoverline{\text{N}}$, respectively.}
\end{figure}

\noindent We need to find the image (or images) of a point object placed at $\text{A}\equiv(a,b)$ in an arbitrary curved mirror defined by equation $y=f(x)$. The mirrors considered are smooth and perfectly reflecting so that refraction, absorption or dispersion of beam itself does not occur. The following shorthand notations will be useful:
\begin{align*}
f&=f(\lambda),\\
f'&=\frac{\ud f}{\ud \lambda},\\
f''&=\frac{\ud^2 f}{\ud \lambda^2}.
\end{align*}
The first task is to find the equation of reflected ray obtained on reflection from a general point on the mirror. Suppose a ray of light from A is incident at a point $\text{P}\equiv(\lambda,f(\lambda))$ on the mirror. The equation of normal at the point of incidence P is,
\begin{equation}
\xoverline{\text{N}}\equiv y-f=-\frac{1}{f'}(x-\lambda).\label{E6}
\end{equation}
The reflected ray $\xoverline{\text{R}}$ passes through P. In order to find its equation we need one more point. Consider the point C on the reflected ray which is symmetrical to A about the normal $\xoverline{\text{N}}$. A line parallel to the tangent $\xoverline{\text{T}}$ at P through A will intersect the normal at B, which in turn is the mid-point of line segment $\xoverline{\text{AC}}$. See Fig. \ref{Fig1}.
The equation of line parallel to tangent through A is given by,
\begin{equation}\label{E5}
\xoverline{\text{I}}\equiv y-b=f'(x-a),
\end{equation}
The simultaneous solution of Eq. \eqref{E6} and \eqref{E5} gives
\begin{align}
f-b&=f'(x_B-a)+\frac{1}{f'}(x_B-\lambda)\nonumber\\
\Rightarrow\left(f'+\frac{1}{f'}\right)x_B&=f-b+af'+\frac{\lambda}{f'}\nonumber\\
\Rightarrow x_B&=\frac{f'}{1+f'^2}\left(f-b+af'+\frac{\lambda}{f'}\right)\nonumber\\
\Rightarrow x_B&=\frac{f'(f-b)}{1+f'^2}+\frac{\lambda}{1+f'^2}+\frac{af'^2}{1+f'^2},\label{xb}
\end{align}
and the corresponding $y$ coordinate is
\begin{align}
y_B&=b+f'\left[\frac{f'(f-b)}{1+f'^2}+\frac{\lambda}{1+f'^2}+\frac{af'^2}{1+f'^2}-a\right]\nonumber\\
\Rightarrow y_B&=\frac{(\lambda-a)f'}{1+f'^2}+\frac{ff'^2}{1+f'^2}+\frac{b}{1+f'^2}.\label{yb}
\end{align}
As stated before, B is the mid-point of A and C, which means
\begin{align}
x_C&=2x_B-a\nonumber\\
\Rightarrow x_C&=2f'\frac{f-b}{1+f'^2}+\frac{2\lambda}{1+f'^2}-\left(\frac{1-f'^2}{1+f'^2}\right)a,\label{E7}
\end{align}
where we insert $x_B$ from Eq. \eqref{xb}. Similarly the $y$ coordinate of C is,
\begin{align}
y_C&=2y_B-b\nonumber\\
\Rightarrow y_C&=2f'\frac{\lambda-a}{1+f'^2}+\frac{2ff'^2}{1+f'^2}+\left(\frac{1-f'^2}{1+f'^2}\right)b, \label{E8}
\end{align}
using expression of $y_B$ from Eq. \eqref{yb}. The equation of the reflected ray is$$\xoverline{\text{R}}\equiv y=mx+c,$$where$$m=m(\lambda)=\frac{y_C-f}{x_C-\lambda},$$and
\begin{equation}
c=c(\lambda)=f-m\lambda.\label{E9}
\end{equation}
The expression of $m(\lambda)$ using Eq. \eqref{E7} and Eq. \eqref{E8} is
\begin{equation}
m(\lambda)=\frac{2(\lambda-a)f'-(f-b)(1-f'^2)}{2(f-b)f'+(\lambda-a)(1-f'^2)}.\label{E10}
\end{equation}
The expression of $c(\lambda)$ using Eq. \eqref{E9} and \eqref{E10} is
\begin{equation}
c(\lambda)=\frac{(2f\lambda-b\lambda-af)(1-f'^2)+2f'(f^2-\lambda^2+a\lambda-bf)}{2(f-b)f'+(\lambda-a)(1-f'^2)}.\label{E11}
\end{equation}
In principle, the parametric equation of the family of reflected rays is obtained. To find the locus of their intersection, proceed as in Sec. \ref{Sec1}. Evaluate $\ud c/\ud m$ as follows:$$\frac{\ud c}{\ud m}=\frac{\ud c/\ud \lambda}{\ud m/\ud \lambda}=\frac{c'}{m'}.$$From Eq. \eqref{E10}, $m'$ is given by
\begin{equation}
D^2m'=\left(1+f'^2\right)^2\left[(f-b)-f'(\lambda-a)\right]+2f''(1+f'^2)\left[(\lambda-a)^2+(f-b)^2\right],
\end{equation}
where for simplicity $$D=2(f-b)f'+(\lambda-a)(1-f'^2).$$From Eq. \eqref{E9} we have the following:$$c'=f'-m-\lambda m',$$so that
\begin{align}
&x(\lambda)=-\frac{\ud c}{\ud m}=\lambda-\frac{f'-m}{m'}\nonumber\\
&\bm{x(\lambda)=\lambda-\frac{\left[(f-b)-f'(\lambda-a)\right]\left[2(f-b)f'+(\lambda-a)(1-f'^2)\right]}{\left(1+f'^2\right)\left[(f-b)-f'(\lambda-a)\right]+2f''\left[(\lambda-a)^2+(f-b)^2\right]}},\label{E12}
\end{align}
and 
\begin{align}
&y(\lambda)=c-m\frac{\ud c}{\ud m}\nonumber\\
&\bm{y(\lambda)=f-\frac{\left[(f-b)-f'(\lambda-a)\right]\left[2(\lambda-a)f'-(f-b)(1-f'^2)\right]}{\left(1+f'^2\right)\left[(f-b)-f'(\lambda-a)\right]+2f''\left[(\lambda-a)^2+(f-b)^2\right]}}.\label{E13}
\end{align}
We have arrived at the Eq. \eqref{E12} and \eqref{E13} which define the curve of image obtained when a point object is placed at $(a,b)$ and the mirror is given by $y=f(x)$. Note that the parameter $\lambda$ is interpreted as the $x$ coordinate varying over the mirror.
\subsection{Application to a few trivial geometries}\label{Trivial}
\noindent Here we consider some cases whose results we know from ray tracing geometric optics \cite{optics}.
\subsubsection*{Reflection in a plane mirror}
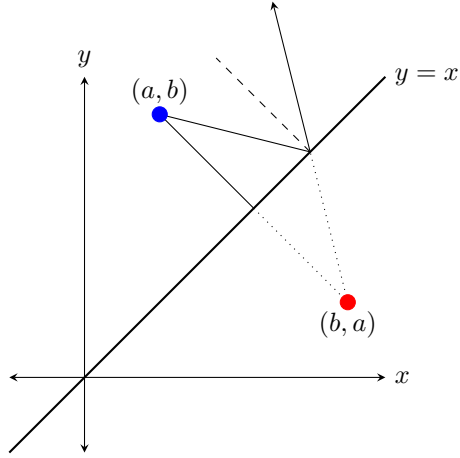
\begin{figure}[t]
\centering
\begin{tikzpicture}[>=stealth]
\draw[<->](-1,0)--(4,0);
\node [right] at (4,0){$x$};
\node [above] at (0,4){$y$};
\draw[<->](0,-1)--(0,4);
\draw[thick](-1,-1)--(4,4);
\node [above] at (1,3.5){$(a,b)$};
\node [below] at (3.5,1){$(b,a)$};
\draw (1,3.5)--(2.25,2.25);
\draw[dotted](2.25,2.25)--(3.5,1);
\draw[->](1,3.5)--(3,3)--(2.5,5);
\draw[dotted](3.5,1)--(3,3);
\draw[dashed](3,3)--(1.75,4.25);
\node [right] at (4,4){$y=x$};
\draw [fill, blue] (1,3.5) circle [radius=0.1];
\draw [fill, red] (3.5,1) circle [radius=0.1];
\end{tikzpicture}
\caption{Image (red dot) of a point object (blue dot) in a plane mirror, $y=x$, shown by a solid black line.}\label{Fig2}
\end{figure}

\noindent Reflection is a plane mirror is perhaps the simplest setup. A common example of a plane mirror is the household looking glass. For plane mirror, the object and its image are symmetrical about the mirror. Let us now see what the Eq. \eqref{E12} and \eqref{E13} give us. The point $\text{A}\equiv(a,b)$ is to be reflected about the line $y=x$. See Fig. \ref{Fig2}.
\begin{align*}
f&=\lambda,\\
f'&=1,\\
f''&=0,
\end{align*}
which give the following:
\begin{align*}
(f-b)-f'(\lambda-a)&=a-b,\\
2(f-b)f'+(\lambda-a)(1-f'^2)&=2(\lambda-b),\\
2(\lambda-a)f'-(f-b)(1-f'^2)&=2(\lambda-a),\\
\left(1+f'^2\right)\left[(f-b)-f'(\lambda-a)\right]&=2(a-b),\\
2f''\left[(\lambda-a)^2+(f-b)^2\right]&=0.
\end{align*}
Putting these into Eq. \eqref{E12} and \eqref{E13} we get$$x(\lambda)=b,\quad y(\lambda)=a.$$Thus, the reflection of a point object placed at $(a,b)$ produces a point image at $(b,a)$ in the plane mirror defined by $y=x$.
\subsubsection*{Reflection in a spherical mirror}
\begin{figure}[t]
    \centering
    \includegraphics[width=1\linewidth]{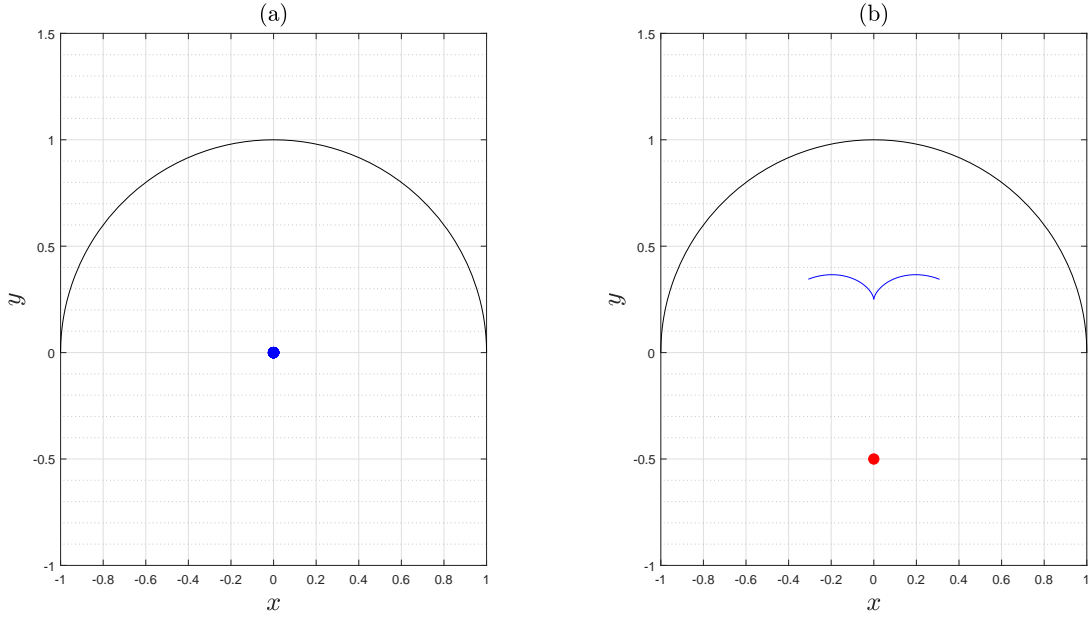}
    \caption{In setup (a), the particle is placed at the centre of spherical mirror while in (b), it is placed at distance $R/2$ below the centre. Plots created on MATLAB for $R=1$ unit.}
    \label{fig:Spherical}
\end{figure}

\noindent A point object is placed at the centre of a circular mirror. Let the equation of circle of radius $R$ be $x^2+y^2=R^2$ and the point object whose image is to be obtained is placed at $\text{A}\equiv(0,0)$. By introducing a parameter $\lambda$ we may define the following,
\begin{align*}
f&=\sqrt{R^2-\lambda^2},\\
f'&=-\frac{\lambda}{f},\\
f''&=-\frac{R^2}{f^3}.
\end{align*}
With these defined, the following will be required in the expression of Eq. \eqref{E12} and \eqref{E13}.
\begin{align*}
(f-b)-f'(\lambda-a)&=\frac{R^2}{f},\\
2(f-b)f'+(\lambda-a)(1-f'^2)&=-\frac{\lambda R^2}{f^2},\\
2(\lambda-a)f'-(f-b)(1-f'^2)&=-\frac{R^2}{f^2},\\
\left(1+f'^2\right)\left[(f-b)-f'(\lambda-a)\right]&=\frac{R^4}{f^3},\\
2f''\left[(\lambda-a)^2+(f-b)^2\right]&=-\frac{2R^4}{f^3}.
\end{align*}
Thus, $$x(\lambda)=0,\quad y(\lambda)=0.$$This is an expected result. Even in Gaussian optics, for the object placed at $u=-2F$ the image is formed on the object itself $v=-2F$ for a concave mirror\footnote{Using the standard sign conventions: object distance being negative, focal length is negative for concave mirror and positive for convex mirror.}. The Fig. \ref{fig:Spherical}(a) illustrates this example. On the other hand if the object is shifted, say, away from the centre to a distance $u=-1.5R=-3F$ from the optical centre (centre of the aperture), then the image does not converge to a unique point. Note that the sharp tip seen in the Fig. \ref{fig:Spherical}(b) image corresponds to the result obtained under the paraxial approximation. The application of the mirror formula (Eq. \eqref{Mirfor}) gives
\begin{align*}
  &\, \frac{1}{v}+\frac{1}{u}=\frac{1}{F}\\
  \Rightarrow &\, \frac{1}{v}+\frac{1}{-3F}=\frac{1}{-F}\\
  \Rightarrow &\, v=-\frac{3F}{2}
\end{align*}
This implies that the image is formed at a distance of $3R/4$ from the aperture centre on the same side as the object.
\subsubsection*{Reflection in a parabolic mirror}
\noindent Next, we can consider another interesting case: a parabolic mirror. Suppose that the parabolic mirror is defined by $y^2=4x$ and the point object is placed at its focus, $\text{F}\equiv(1,0)$.
\begin{align*}
f&=2\sqrt{\lambda},\\
f'&=\frac{1}{\sqrt{\lambda}},\\
f''&=-\frac{1}{2\lambda^{3/2}},
\end{align*}
which give the following:
\begin{align*}
(f-b)-f'(\lambda-a)&=\sqrt{\lambda}+\frac{1}{\sqrt{\lambda}},\\
2(f-b)f'+(\lambda-a)(1-f'^2)&=\frac{(\lambda+1)^2}{\lambda},\\
2(\lambda-a)f'-(f-b)(1-f'^2)&=0,\\
\left(1+f'^2\right)\left[(f-b)-f'(\lambda-a)\right]&=\frac{(\lambda+1)^2}{\lambda^{3/2}},\\
2f''\left[(\lambda-a)^2+(f-b)^2\right]&=-\frac{(\lambda+1)^2}{\lambda^{3/2}}.
\end{align*}
On substituting these into Eq. \eqref{E12} and \eqref{E13}, both $x(\lambda)$ and $y(\lambda)$ become undefined. This is not surprising because we know from the properties of parabola, that the normal at a point on the parabola bisects the angle between the line joining the focus and the line parallel to the principle axis \cite{loney}. In other words, rays coming in parallel to the principle axis of a parabola meet at its focus.
\begin{figure}[t]
\centering
\begin{tikzpicture}[>=stealth]
\draw[<->](-1,0)--(4,0);
\draw[<->](0,-4)--(0,4);
\node [right] at (4,0){$x$};
\node [above] at (0,4){$y$};
\draw[thick, domain=0:0.4] plot (\x, {2*sqrt(\x)});
\draw[thick, domain=0.4:4] plot (\x, {2*sqrt(\x)});
\draw[thick, domain=0:0.4] plot (\x, {-2*sqrt(\x)});
\draw[thick, domain=0.4:4] plot (\x, {-2*sqrt(\x)});
\draw[->, thick](1,0)--(0.25,1)--(4,1);
\draw[->, thick](1,0)--(0.25,-1)--(4,-1);
\draw[->, thick](1,0)--(2,2.828)--(4,2.828);
\draw[->, thick](1,0)--(2,-2.828)--(4,-2.828);
\node [above right] at (1.1,0.1){F};
\node [above] at (1,2.5){$y^2=4ax$};
\node [left] at (4,1.7){Meet at $\infty$};
\node [left] at (4,-1.7){Meet at $\infty$};
\draw [fill, blue] (1,0) circle [radius=0.1];
\end{tikzpicture}
\caption{The point object (blue dot) is placed at the focus of the parabola. The image at formed at infinity.}
\end{figure}
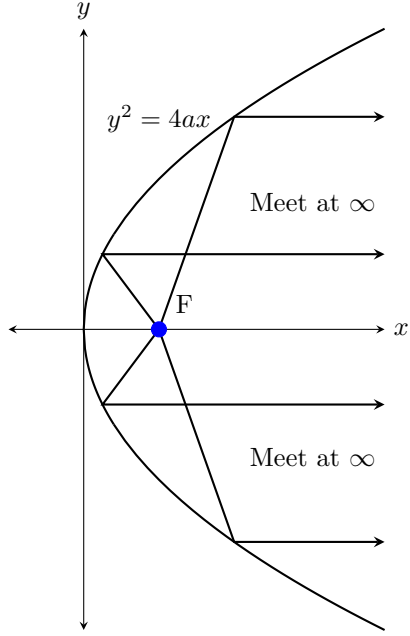
\subsubsection*{Reflection in an elliptic mirror}
\begin{figure}[t]
    \centering
    \includegraphics[width=1\linewidth]{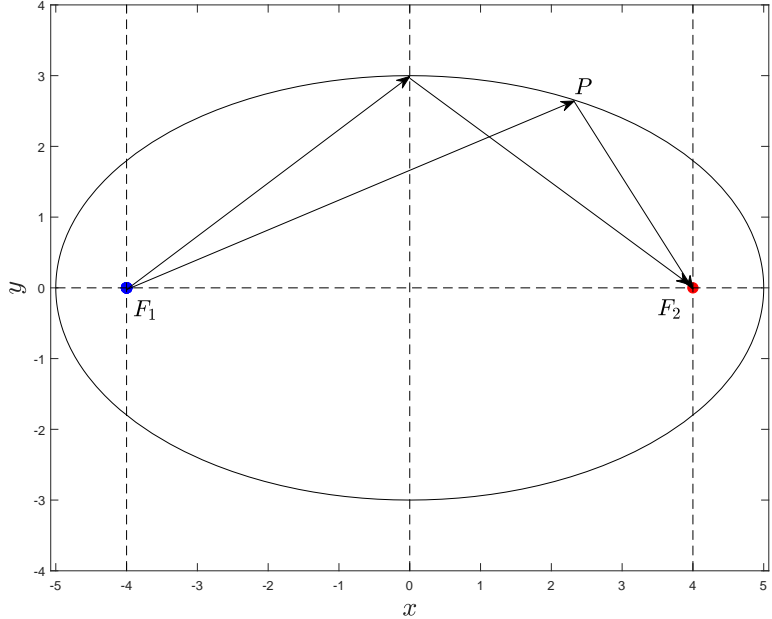}
\caption{Reflection of one focus from the inner surface of the ellipse. This ellipse chosen for illustration has $A=5,B=3,\textup{e}=0.8$ so that the foci are located at $(\pm A\textup{e},0)=(\pm4,0)$.}\label{Ellip}
\end{figure}

\noindent Yet another simple example is of reflection from the inner surface of an ellipse. Where is the image formed when a point object is placed at one of the focus of the ellipse? For this consider an ellipse centred at the origin with its axis parallel to the coordinate axis. The equation of such an ellipse \cite{loney} would be $$\frac{x^2}{A^2}+\frac{y^2}{B^2}=1,$$where $A$ and $B(<A)$ are the semi-major and semi-minor axis, respectively. The lengths $A$ and $B$ are related by the eccentricity e as
\begin{equation}
    B^2=A^2(1-\text{e}^2).\label{Rel_AB}
\end{equation}
For$$\frac{\lambda^2}{A^2}+\frac{f^2}{B^2}=1,$$we have
\begin{align*}
f'&=-\frac{B^2}{A^2}\frac{\lambda}{f},\\
f''&=-\frac{B^4}{A^2f^3}.
\end{align*}
Suppose that the point object is placed at $\text{F}_1\equiv(-A\text{e},0)$ then
\begin{align*}
(f-b)-f'(\lambda-a)&=\frac{B^2}{Af}(A-\lambda\text{e}),\\
2(f-b)f'+(\lambda-a)(1-f'^2)&=-\frac{B^2}{A^2f^2}(\lambda+A\text{e})(A-\lambda\text{e})^2,\\
2(\lambda-a)f'-(f-b)(1-f'^2)&=-\frac{B^2}{A^2f}(A-\lambda\text{e})^2,\\
\left(1+f'^2\right)\left[(f-b)-f'(\lambda-a)\right]+2f''\left[(\lambda-a)^2+(f-b)^2\right]&=\frac{(\lambda\text{e}-A)^3B^4}{A^3f^3}.
\end{align*}
Using Eq. \eqref{E12}, \eqref{E13} and \eqref{Rel_AB} we arrive at $$(x(\lambda),y(\lambda))\equiv(A\text{e},0).$$Thus, the image of a point placed at one of the foci is formed at the other focus. This could have been directly predicted from the properties of ellipse. The normal at any point P on the ellipse bisects the angle $\text{F}_1\text{PF}_2$, where F$_1$ and F$_2$ are the foci of the ellipse \cite{loney}. See Fig. \ref{Ellip}.
\subsection{Some more examples}
For the examples considered in this section it will be easier to analyse equations \eqref{E12} and \eqref{E13} only graphically.
\subsubsection*{Reflection in ``cubic'' mirror}
\begin{figure}[t]
    \centering
    \includegraphics[width=1\linewidth]{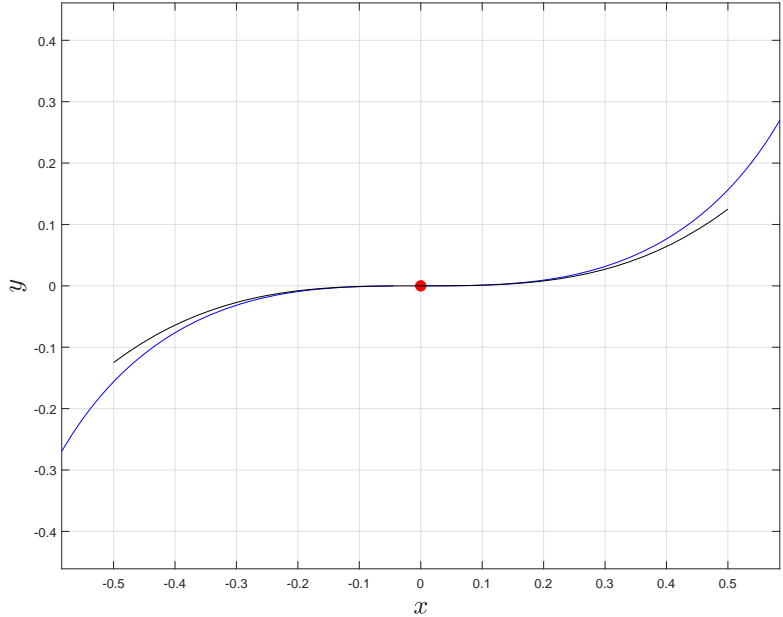}
    \caption{The point object (red dot) is placed at the origin. The black solid line is the mirror shaped like $y=x^3$ and the blue solid line is the image curve. The mirror extends for $x\in[0.5,0.5]$ .}
    \label{x3}
\end{figure}

\noindent Let the mirror be in the shape of $y=x^3$ and the point object be located at $(0,0)$. On applying Eq. \eqref{E12} and \eqref{E13}, we will get
\begin{align*}
    x(\lambda)&=6\lambda\left(\frac{1-\lambda^4}{5-3\lambda^4}\right)\\
    y(\lambda)&=2\lambda^3\left(\frac{5+3\lambda^4}{5-3\lambda^4}\right)
\end{align*}
Note that because the object lies on the mirror itself so does its image, atleast near the object but a deviation can be clearly seen (Fig. \ref{x3}) farther away from the object. The diagram shown is exactly for the range $\lambda\in\left[-\frac{1}{2}, \frac{1}{2}\right]$.
\subsubsection*{Reflection in an exponentially shaped mirror}
\begin{figure}[t]
\centering
\includegraphics[width=1\linewidth]{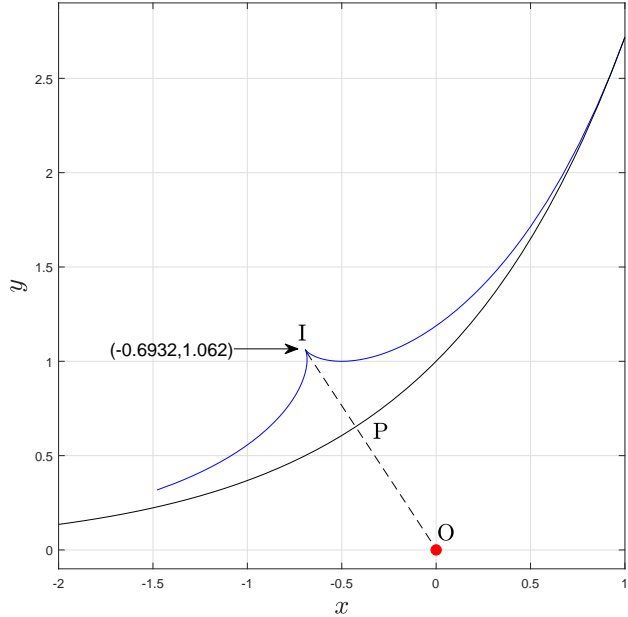}
\cprotect\caption{Reflection of origin O (red dot) in $y=e^x$ (solid black line). The extent of the image (shown by blue line) corresponds exactly to the aperture of the mirror considered, i.e. for $\lambda\in[-2,1]$. The black dashed line is the optical axis of the setup if paraxial approximation is considered. The point `I' is marked using the \verb|Data Cursor| feature of MATLAB.}\label{expo}
\end{figure}

\noindent Consider the image of a point object located at the origin in the mirror $y=e^x$. On putting the Eq. \eqref{E12} and \eqref{E13} in a plotting software (such as MATLAB) the Fig. \ref{expo} is obtained. To apply the Gaussian optics for this setup we would first need to locate the optical axis of this system so that the rays are paraxial. This corresponds to finding the normal to the mirror which passes through the object. Suppose the normal at $\text{P}\equiv(x_0,e^{x_0})$ to $y=e^x$ passes through the origin. Then the following condition must be satisfied,
\begin{align*}
    \frac{e^{x_0}-0}{x_0-0}&=-\left.\frac{\ud x}{\ud y}\right|_{x_0}\\
    e^{2x_0}+x_0&=0.
\end{align*}
The above equation may be solved computationally to obtain $x_0=-0.4263$. Hence the object distance is given by$$u=-\sqrt{x_0^2+e^{2x_0}}=-0.7798$$The radius of an equivalent spherical mirror at the point P can be approximated by the radius of curvature of the curve at P. The general formula for the radius of curvature is \cite{Thomas} $$R=\frac{(1+y'^2)^{3/2}}{y''}$$For $y=e^x$ at P it is$$R=\left.\frac{(1+y'^2)^{3/2}}{y''}\right|_{x_0}=2.6089,$$so that the focal length is$$F=\frac{R}{2}=1.3045.$$The application of mirror formula Eq. \eqref{Mirfor} gives
\begin{align*}
    \frac{1}{v}&+\frac{1}{-0.7798}=\frac{1}{1.3045}\\
    v&=0.4880.
\end{align*}
The positive sign of $v$ implies that the image is formed on the side opposite to the object. The coordinates of the sharp tip (labelled as I) in the image curve are  $(-0.6932,1.062)$. The distance IP is given by
\begin{align*}
    \text{IP}&=\sqrt{(x_0+0.6932)^2+(e^{x_0}-1.062)^2}\\
    \text{IP}&=0.4884.
\end{align*}
The image distance obtained using the mirror formula is $v=0.4880$ which is comparable to the exact value IP$=0.4884$ obtained from Eq. \eqref{E12} and \eqref{E13}. Note the limitation of mirror formula which can only predict point I as the image of O.
\section{Limiting Case}\label{LC}
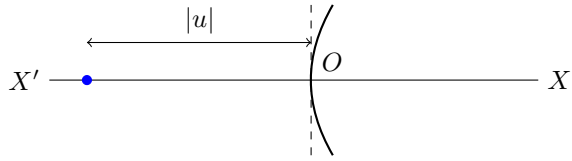
\begin{figure}[t]
\centering
\begin{tikzpicture}
\draw[thick] (0.269,1) to [out=240, in= 120] (0.269,-1);
\draw[<->](-3,0.5)--(-0.02,0.5);
\draw[dashed](-0.02,-1)--(-0.02,1);
\node [above] at (-1.5,0.5){$|u|$};
\node [above right] at (0,0){$O$};
\draw(-3.5,0)--(3,0);
\node [right] at (3,0){$X$};
\node [left] at (-3.5,0){$X'$};
\draw [fill, blue] (-3,0) circle [radius=0.06];
\end{tikzpicture}
\caption{The blue dot represents the object, $X'OX$ is the optical axis and $O$ is the optical centre.}\label{Para}
\end{figure}

\noindent In this section we will prove that the limiting case of Eq. \eqref{E12} and \eqref{E13} under the paraxial approximation give back the mirror formula \eqref{Mirfor} for spherical mirrors. Without the loss of generality, assume the spherical mirror of radius $|R|$ is centred at $(R,0)$ and the point object is placed at $(u,0)$. The $R$ can be negative as well, in which case the object would be facing the concave side of the mirror. See Fig. \ref{Para}. The equation of circle is $(\lambda-R)^2+f^2=R^2$ or $f^2=\lambda(2R-\lambda)$ so that
\begin{align*}
f'&=\frac{R-\lambda}{f},\\
f''&=-\frac{R^2}{f^3},
\end{align*}
which give the following
\begin{align*}
(f-b)-f'(\lambda-a)&=\frac{1}{f}(-uR+\lambda R+u\lambda),\\
2(f-b)f'+(\lambda-a)(1-f'^2)&=\frac{1}{f^2}\left[(2R-u-\lambda)f^2-(\lambda-u)(\lambda-R)^2\right],\\
2(\lambda-a)f'-(f-b)(1-f'^2)&=\frac{1}{f}(R^2-2R\lambda-2uR+2u\lambda),\\
\left(1+f'^2\right)\left[(f-b)-f'(\lambda-a)\right]+2f''\left[(\lambda-a)^2+(f-b)^2\right]&=-\frac{R^2}{f^3}\left[(\lambda-u)(\lambda+R-2u)+f^2\right].
\end{align*}
On inserting the above expressions into Eq. \eqref{E12} and \eqref{E13} we get,
\begin{align}
x&=\lambda-\frac{(-uR+\lambda R+u\lambda)}{R^2}\left[\frac{(2R-u-\lambda)f^2-(\lambda-u)(\lambda-R)^2}{(\lambda-u)(\lambda+R-2u)+f^2}\right]\label{E21}\\
y&=f-f\frac{(-uR+\lambda R+u\lambda)}{R^2}\left[\frac{R^2-2R\lambda-2uR+2u\lambda}{(\lambda-u)(\lambda+R-2u)+f^2}\right]\label{E22}
\end{align}
Now we can make use of the paraxial approximation where the aperture or the extent of the mirror considered is small. In terms of variables, we may take$$\lambda\approx0 \text{ and }f\approx 0$$because light rays considered are close to the optical axis $X'OX$, where $O$ is the optical centre of the system. Putting $\lambda,f=0$ in Eq. \eqref{E21} and \eqref{E22}, they reduce to
\begin{align}
&x=-\frac{Ru}{R-2u}\nonumber\\
&\frac{1}{x}+\frac{1}{u}=\frac{2}{R}\label{E23}
\end{align}
and $y=0$, respectively. Hence, the location of the image given by the $x$ coordinate in Eq. \eqref{E23}. Because the object considered is a point lying on the optical axis, the $y$ coordinate not surprisingly is 0.
\section{Conclusion}
\noindent We considered the image of a point object in a curved mirror defined by an equation on the Cartesian plane. The main equations of this work are \eqref{E12} and \eqref{E13}. These equation give back the standard mirror formula of Gaussian optics as proved in Sec. \ref{LC} for appropriate approximations. The work done in this paper may be considered for extended objects. One could go even further and repeat the calculations in higher dimensions. They will require deeper analysis and longer calculations and will appear in a separate publications. 
\bibliographystyle{unsrt}
\bibliography{Biblo}
\vfill
\end{document}